\documentclass[%
 twocolumn,
superscriptaddress,
 amsmath,amssymb,
https://www.overleaf.com/1538519932mmpfzbyxcfdc prb,
https://www.overleaf.com/1538519932mmpfzbyxcfdc 
floatfix,
citeautoscript,
]{revtex4-2}

\usepackage{graphicx}
\usepackage{dcolumn}
\usepackage{bm}
\usepackage{hyperref}

\usepackage[scale=0.794]{geometry}
\usepackage{xcolor}
\usepackage[final]{changes}
\definechangesauthor[color=blue]{common}
\definechangesauthor[color=red, name={Paulo Santos}]{PVS}
\definechangesauthor[color=green]{MM}
\setauthormarkup{}

\newcommand{\rPVS}[2]{{#1}{}}
\newcommand{\aPVS}[1]{{#1}}
\newcommand{\dPVS}[1]{{}}

\newcommand{\rPVSn}[2]{\replaced[id=PVS]{#1}{#2}}
\newcommand{\aPVSn}[1]{\added[id=PVS]{#1}}

\begin{document}



\preprint{LBAW-GaAs}

\title{Generation and propagation of super-high-frequency
\\bulk acoustic wave in GaAs }


\author{Diego H. O. Machado}
\affiliation{Paul-Drude-Institut f{\"u}r Festk{\"o}rperelektronik, Leibniz-Institut im Forschungsverbund Berlin e. V., Hausvogteiplatz 5-7, 10117 Berlin, Germany }
\affiliation{UNESP, S{\~a}o Paulo State University, Department of Physics, FC and Graduate Program in Materials Science and Technology (POSMAT), \\Av. Eng. Luiz Edmundo C. Coube 14-01, 17033-360, Bauru, SP, Brazil.}
 
\author{Antonio Crespo-Poveda}
\affiliation{Paul-Drude-Institut f{\"u}r Festk{\"o}rperelektronik, Leibniz-Institut im Forschungsverbund Berlin e. V., Hausvogteiplatz 5-7, 10117 Berlin, Germany }

\author{Alexander S. Kuznetsov}
\affiliation{Paul-Drude-Institut f{\"u}r Festk{\"o}rperelektronik, Leibniz-Institut im Forschungsverbund Berlin e. V., Hausvogteiplatz 5-7, 10117 Berlin, Germany }

\author{Klaus Biermann}
\affiliation{Paul-Drude-Institut f{\"u}r Festk{\"o}rperelektronik, Leibniz-Institut im Forschungsverbund Berlin e. V., Hausvogteiplatz 5-7, 10117 Berlin, Germany }

\author{Luis V. A. Scalvi}
\affiliation{UNESP, S{\~a}o Paulo State University, Department of Physics, FC and Graduate Program in Materials Science and Technology (POSMAT), \\Av. Eng. Luiz Edmundo C. Coube 14-01, 17033-360, Bauru, SP, Brazil.}

\author{Paulo V. Santos}
\affiliation{Paul-Drude-Institut f{\"u}r Festk{\"o}rperelektronik, Leibniz-Institut im Forschungsverbund Berlin e. V., Hausvogteiplatz 5-7, 10117 Berlin, Germany }
 \email{santos@pdi-berlin.de}

\date{\today}

\begin{abstract}
Coherent super-high-frequency (SHF) vibrations provide an excellent tool for the modulation and control of excitation\aPVSn{s} in semiconductors. Here, we investigate the piezoelectric generation and propagation of longitudinal bulk acoustic waves (LBAWs) with frequencies up to 20 GHz in GaAs crystals using bulk acoustic wave resonators (BAWRs) based on piezoelectric thin ZnO films. We show that the electro-acoustic conversion efficiency of the BAWRs depends sensitively on the sputtering conditions of the ZnO films. The BAWRs were then used for the study of the propagation properties of the LBAWs in GaAs in the frequency and temperature ranges from 1 to 20 GHz  and $10$ and $300$ K, respectively, which have so far not been experimentally accessed. We found that the \rPVS{acoustic absorption}{~intrinsic acoustic absorption $\alpha_\mathrm{eff}$} of GaAs in the temperature range from 80 K to 300 K is dominated by scattering with thermal phonons. At lower temperatures, in contrast, \rPVS{the acoustic absorption}{~$\alpha_\mathrm{eff}$} saturates at a frequency-dependent value. Experiments carried out with different propagation lengths indicate that the saturation is associated with losses during reflections at the sample boundaries. We also demonstrate devices with high quality factor fabricated on top of acoustic Bragg-reflectors. The results presented here prove the feasibility of high-quality acoustic resonators embedding GaAs-based nanostructures, thus opening the way for the modulation and control of their properties by electrically excited SHF LBAWs. 

\end{abstract}

\maketitle



\section{\label{sec:Introduction} Introduction}

\begin{figure}[t!]
    \centering
    \includegraphics[width=1.01\columnwidth, angle=0, clip]{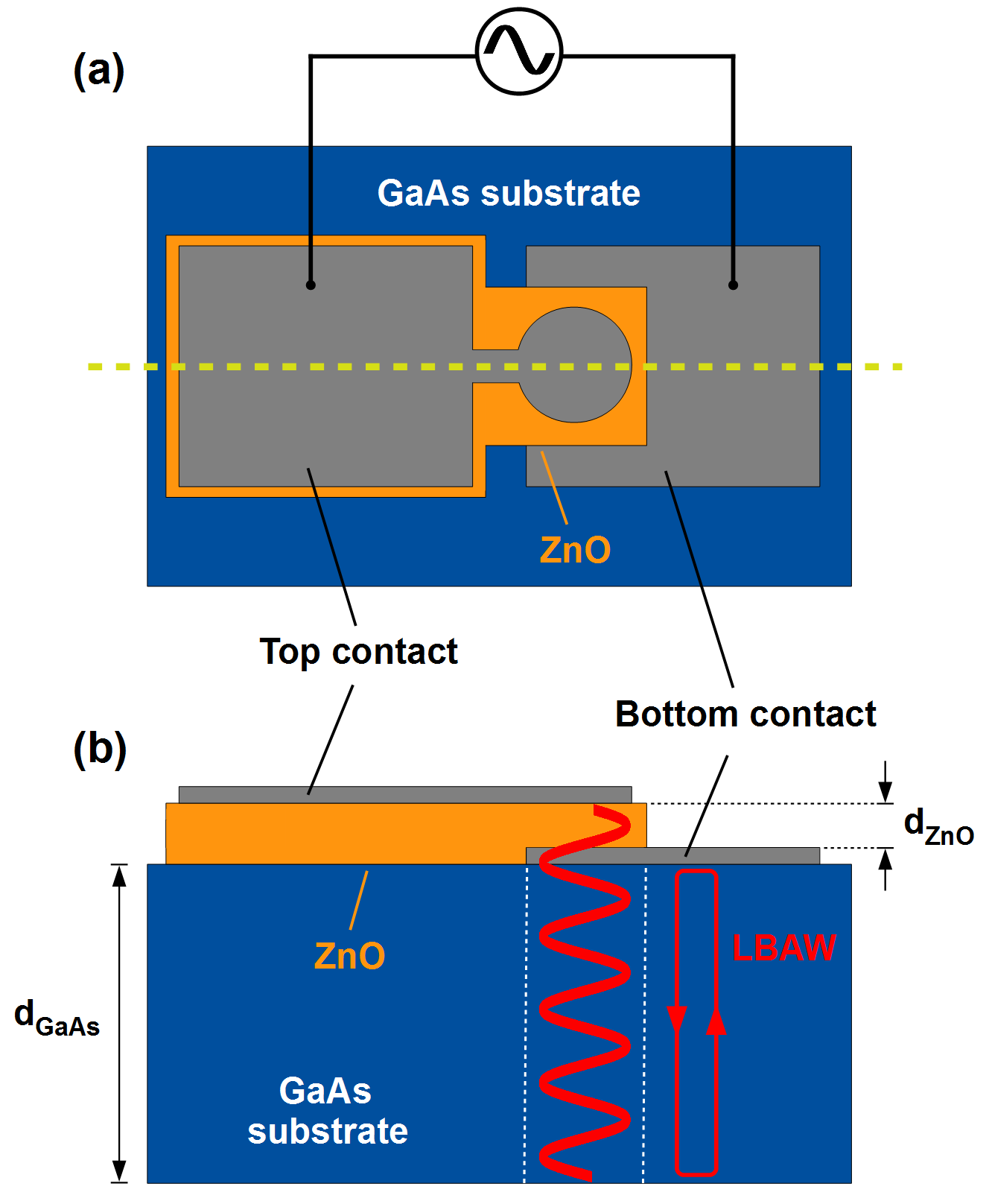}
    \caption{Bulk acoustic wave resonator (BAWR) for the generation of super-high-frequency longitudinal bulk acoustic waves (LBAW) based on a piezo-electric ZnO thin film of thickness $d_{\mathrm{ZnO}}$: (a) Top view showing the metal contacts.  (b) Cross-section along the horizonal dashed line in (a). LBAW generation takes place in the overlapping area of the metal contacts, which is  delimited by the vertical dashed lines. $d_{\mathrm{GaAs}}$ denotes the thickness of the GaAs substrate.}
    \label{Figure1}
\end{figure}

Acousto-electric devices find applications in different areas, including signal processing, sensorics, and acousto-electric modulation. Several recent studies have demonstrated that acoustic waves are an excellent tool for the manipulation and control of opto-electronic excitation\aPVSn{s} in these structures (see Ref.~\cite{PVS326} for a recent review). Both surface and bulk acoustic vibrations can be electrically excited by applying a microwave electromagnetic field to a piezoelectric material. Surface acoustic waves (SAWs) are normally generated by applying the field via interdigitated transducers deposited on a piezoelectric surface~\cite{White70a}. Bulk acoustic waves (BAWs) can be conveniently generated by bulk acoustic wave resonators (BAWRs) consisting of a piezoelectric layer sandwiched between two metallic contacts, as illustrated in Fig.~\ref{Figure1}(a)~\cite{Auld90a, Royer00a}.

The \rPVSn{last two}{~two last} decades have witnessed significant efforts to increase the operation frequency of acousto-electric devices to the super-high-frequency (SHF) range extending from 3 to 30 GHz. Here, one of the motivations has been the opening of new high-frequency bands for mobile telecommunication generating the demand for filters in the few GHz range. Operation at high frequencies also increases the sensitivity of acousto-electric sensors and improves the time response of opto-mechanical structures as well as acousto-optic modulators. For the above applications, the leakage of the vibration induced in the piezoelectric layer into the substrate is undesirable. Therefore, the BAWR are normally fabricated on a suspended membrane (these structures are normally known as thin film bulk acoustic resonators, FBARs) or on a layer stack designed as an acoustic reflector, thus forming a solidly mounted resonator (SMR). 

The availability of electrically excited and coherent SHF vibrations also open\aPVSn{s} the way for novel functionalities. Vibrations interact with a wide range of electronic and magnetic transitions in solid matter: some of these transitions with present technological relevance lie in the SHF range. Examples are the quantum levels of nanomechanical oscillators~\cite{Aspelmeyer_rmp86_1391_14,Oconnell_n464_697_10}, superconducting quantum bits~\cite{PVS241,Satzinger_N563_661_18}, quantum dots~\cite{Naber_PRL96_136807_06,Metcalfe_PRL105_37401_10} and two-level single-photon emission centers~\cite{PVS218,Golter_PRL116_143602_16}, and magnetic excitations~\cite{Weiler_PRL106_117601_11}. Furthermore, vibrations with frequencies above 3 GHz have equivalent temperatures exceeding the temperatures presently achieved in low-temperature refrigerators (down to 10 mK), thus opening a way to isolate and manipulate single vibration quanta. 

Several groups have demonstrated the piezoelectric generation of longitudinal bulk acoustic waves (LBAWs) with frequencies between 10~GHz and 20~GHz~\cite{Rinaldi_ITUFFC57_38_10,Loebl_MCP79_143_03,Lanz_isssus_03,Umeda_V83_672_08}. In most of these cases, the structures consist of AlN-based \rPVSn{}{suspended} FBARs or SMRs on acoustic reflectors fabricated on a silicon or diamond substrate. In parallel to the developments on BAWRs, the piezo-electrical excitation of \rPVSn{SAWs}{~surface acoustic waves (SAW)} has been extended to frequencies up to 32 GHz in GaAs \aPVS{crystals}~\cite{Kukushkin_APL85_4526_04} and \rPVS{up to 24~GHz in silicon crystals coated by a piezoelectric ZnO film}{~24~GHz in silicon}~\cite{PVS263}. 

The optoelectronic properties together with the mature technology for the growth of high-quality nanostructures make (In,Al,Ga)As alloys a relevant material-system for nano-mechanical science and technology. Consequently, techniques for the generation of LBAWs in these materials as well as the precise knowledge of their propagation properties become  key issues for the realization of acoustic resonators with a high quality factor Q. Despite the technological significance of (In,Al,Ga)As alloys, experimental data on the propagation of SHF acoustic waves in GaAs remain very scarce. In particular, there is gap in the experimental determination of the acoustic absorption between the sub-GHz range probed by ultrasonic propagation~\cite{Cottam_JPCSSP7_2447_74} and the frequency range starting at about 50 GHz~\cite{Legrand_PRB93_184304_16,Ivanov_SPSS13_508_71,Chen_PMB70_687_94,Maznev_APL102_41901_13}, which is normally accessed using laser acoustics (see, for instance, the compilation of experimental results summarized in Fig.~3 of Ref.~\cite{Maznev_APL102_41901_13}).


In this paper, we investigate the electric generation and propagation of SHF LBAWs in GaAs crystals using BAWRs of the type illustrated in Fig.~\ref{Figure1}(a). 
\aPVS{These structures require a piezoelectric thin film:  as previously mentioned, both textured ZnO and AlN can be used for this purpose. AlN is better suited for high frequency BAWRs due to the higher acoustic velocity and compatibility with the silicon CMOS technology~\cite{PVS252}. The elastic properties of ZnO, in contrast, are better matched to the ones of (Al,Ga)As alloys, which is a favorable property in particular for SAW applications. In the initial stages of this work, we have fabricated  BAWRs using both textured AlN \protect\footnote{D. Machado {\it et al.}, private commun.} and ZnO layers deposited by sputtering. We found that both types of films lead to  comparable BAWR performances up to frequencies of at least 10 GHz. In the present studies, we will concentrate on BAWRs fabricated using sputtered ZnO films.
}
The remaining sections of the manuscript are organized as follows. The procedures used for the fabrication of these BAWRs, including the fabrication steps for the deposition of contacts sandwiching the piezoelectric ZnO layers, are described in Sec.~\ref{Experimental_details}. The experimental results are presented in Sec.~\ref{Results}. Here, we first address the deposition of high-quality piezoelectric layers (Sec.~\ref{sec:Sec3_1}) and then proceed to the analysis of the electrical response of SHF BAWRs deposited on bare GaAs substrates (Sec.~\ref{Sec3_2}) as a function of frequency and temperature (Sec.~\ref{Sec3_4}). These studies are then extended to structures fabricated on acoustic Bragg reflectors grown epitaxially on GaAs substrates in Sec.~\ref{Sec3_5}. The BAWRs are then employed for the study of LBAW propagation in GaAs over a wide range of frequencies and temperatures. Section~\ref{Discussions} is devoted to the mechanisms responsible for the acoustic absorption. Here, we first identify the different mechanisms leading to acoustic absorption (Sec.~\ref{Sec4_1}). Their behavior at room and cryogenic temperatures is then addressed in Secs.~\ref{Sec4_2} and~\ref{Sec4_3}, respectively. Finally, the main results of the paper are summarized in Sec.~\ref{Conclusions}.


\section{\label{Experimental_details} Experimental details}

\subsection{ \label{sec:BAW_transducer_design} BAW transducer design}

\begin{figure*}[htpb]
    \centering
    \includegraphics[width=1.90\columnwidth, angle=0, clip]{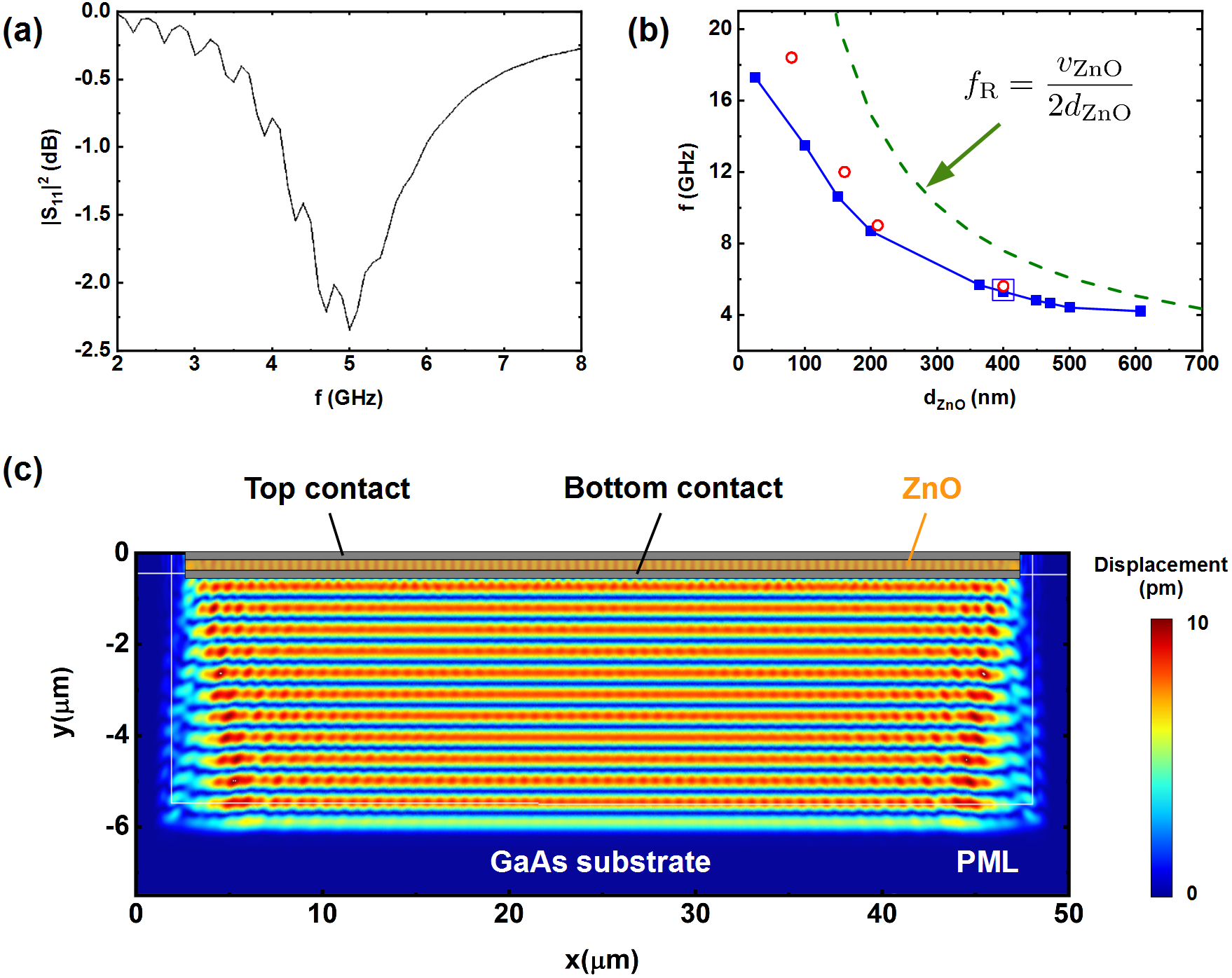}
    \caption{(a) Calculated dependence of the \rPVSn{$|s_\mathrm{11}|^2$ scattering parameter (corresponding to the rf-reflection coefficient)}{~$s_\mathrm{11}$ rf-reflection coefficient} for a BAWR on a GaAs (001) wafer consisting of a $d_\mathrm{ZnO}=440$~nm-thick ZnO layer sandwiched between bottom and top metal contacts formed by a 40 nm Au film and a 10 nm/30 nm/10 nm Ti/Al/Ti layer stack, respectively. The active area of the resonator is equal to $1960~\mu$m$^2$. (b) Dependence of the resonance frequency on $d_\mathrm{ZnO}$: the joined full square symbols are finite element method calculations, and the open circle symbols are experimental values. The dashed line reproduces the approximation for the resonance frequency given by Eq.~(\ref{EqfR}). The blue open square at $f_{R}\cong 5$~GHz corresponds to the device simulated in (a) and (c). (c) Cross-section map of the amplitude of the displacement field of the BAWR in (a) at the resonance frequency. 
    The calculation domain was surrounded by perfectly matched layers (PMLs) to reduce acoustic reflections from the sample boundaries.}
    \label{Figure2}
\end{figure*}

To a first approximation, the central resonance frequency of the BAWRs of Fig.~\ref{Figure1} is given by: 
\begin{equation}
f_R\cong \frac{v_\mathrm{ZnO}}{2d_\mathrm{ZnO}},
\label{EqfR}
\end{equation} 

\noindent where $v_\mathrm{ZnO}=6070$~m/s is the propagation velocity of the LBAWs in the ZnO film\aPVS{~\cite{Gualtieri_ITUFFC41_53_94}} and $d_\mathrm{ZnO}$ its thickness. The previous approximation for the resonance frequency neglects the influence of the metal contacts on the resonant modes and, thus, strictly applies only for structures with thick piezoelectric layers. In order to obtain more reliable results, we have carried out numerical simulations of the frequency dependence of the excited BAWs using a finite element method~\cite{getdp-ieee1998,PVS316}. The simulations were performed by exciting the BAWR with \aPVS{radio frequency (rf)} voltages of different frequencies and solving the coupled piezoelectric and mechanical equations to determine the spatial distribution of the acoustic field as well as the electric power converted to the acoustic modes. In the calculations, we used the lateral dimensions as well as the nominal thicknesses of the ZnO layers and metal contacts of the fabricated samples (see caption of Fig.~\ref{Figure2}). The thickness of the GaAs substrate, in contrast, was set to $5~\mu$m to limit the calculation time. In order to minimize acoustic reflections, the lateral and bottom sides of the calculation domain were surrounded by perfectly matched layers (PMLs).

Figure~\ref{Figure2}(a) displays the \rPVS{rf}{~radio frequency (rf)} scattering parameter $s_\mathrm{11}$ calculated for a BAWR with a 440 nm thick ZnO layer and an area of $1960~\mu$m$^2$ (corresponding to a circular top contact diameter of $25~\mu$m). $s_\mathrm{11}$ quantifies the fraction of the electric power reflected by the BAWR. The pronounced dip centered around the resonance frequency $f_R=5$~GHz arises from the reduced electric reflection as the electrical input drive is converted to an acoustic mode. The oscillations superimposed on the curve are interference fringes due to residual reflections at the back PML. Figure~\ref{Figure2}(b) displays the calculated (full square symbols) as well as the measured (open circle symbols, see Sec.~\ref{Sec3_3} for\rPVSn{}{a} details) dependence of $f_R$ on the thickness of the ZnO film. The dashed line reproduces the simple first-order approximation for the resonance frequency obtained from Eq.~(\ref{EqfR}). This approximation considerably overestimates the resonance frequency, in particular for structures with thin ZnO layers and, thus,  high resonance frequencies. Finally, Fig.~\ref{Figure2}(c) shows the spatial distribution of the acoustic displacement field $|u|$ in the resonator in Fig.~\ref{Figure2}(a) as determined for a fixed phase of the applied rf-voltage. The stripes on the cross-section of the device correspond to the acoustic wave fronts, which are separated by the half of the LBAW period.  

\subsection{BAWR fabrication}

The BAWRs of Fig.~\ref{Figure1}(a) were deposited on nominally intrinsic GaAs (001) wafers (Wafer Technology Ltd.) with a resistivity $> 5.9\times10^7~\Omega\cdot$cm and Hall mobility $> 5100~\mathrm{cm}^2/$(V$\cdot$s). The wafers are double-side polished with a thickness $d_\mathrm{GaAs}=(356 \pm 8)~\mu$m. The bottom contact of BAWRs consists of a 30 nm-thick Au film deposited on a 10-nm-thick Ti adhesion layer. The metal films were deposited via electron-beam evaporation while the ZnO film was fabricated by rf magnetron-sputtering.

The piezoelectric properties of the textured ZnO thin films rely on the preferential orientation of the hexagonal c-axis along the direction perpendicular to the surface, which can be controlled by a proper selection of the sputtering conditions and underlying substrate. Au was employed as the back contact due to its favorable impact on the nucleation of the textured ZnO films, a necessary condition for high piezoelectric coupling~\cite{Wagers_IEEEUS72_194_72,PVS309}. The sputtering process was carried out in a high-vacuum chamber (base pressure $<10^{- 6}$~mbar) with a 5-inch ZnO target \rPVSn{placed 8~cm above the sample}{~displaced from the samples by 8~cm}. The sputtering atmosphere consists of an $80:20$ Ar/O$_2$ gas mixture (pressure of $5\times 10^{-3}$ mbar). The extra O$_2$ atmosphere ensures the formation of high resistivity, stoichiometric films. The rf sputtering powers were typically of about 100~W. As will be discussed in detail below, the structural and piezoelectric properties of the ZnO films are very sensitive to the sputtering conditions.

\begin{figure}[t!]
    \centering
    \includegraphics[width=1.01\columnwidth, angle=0, clip]{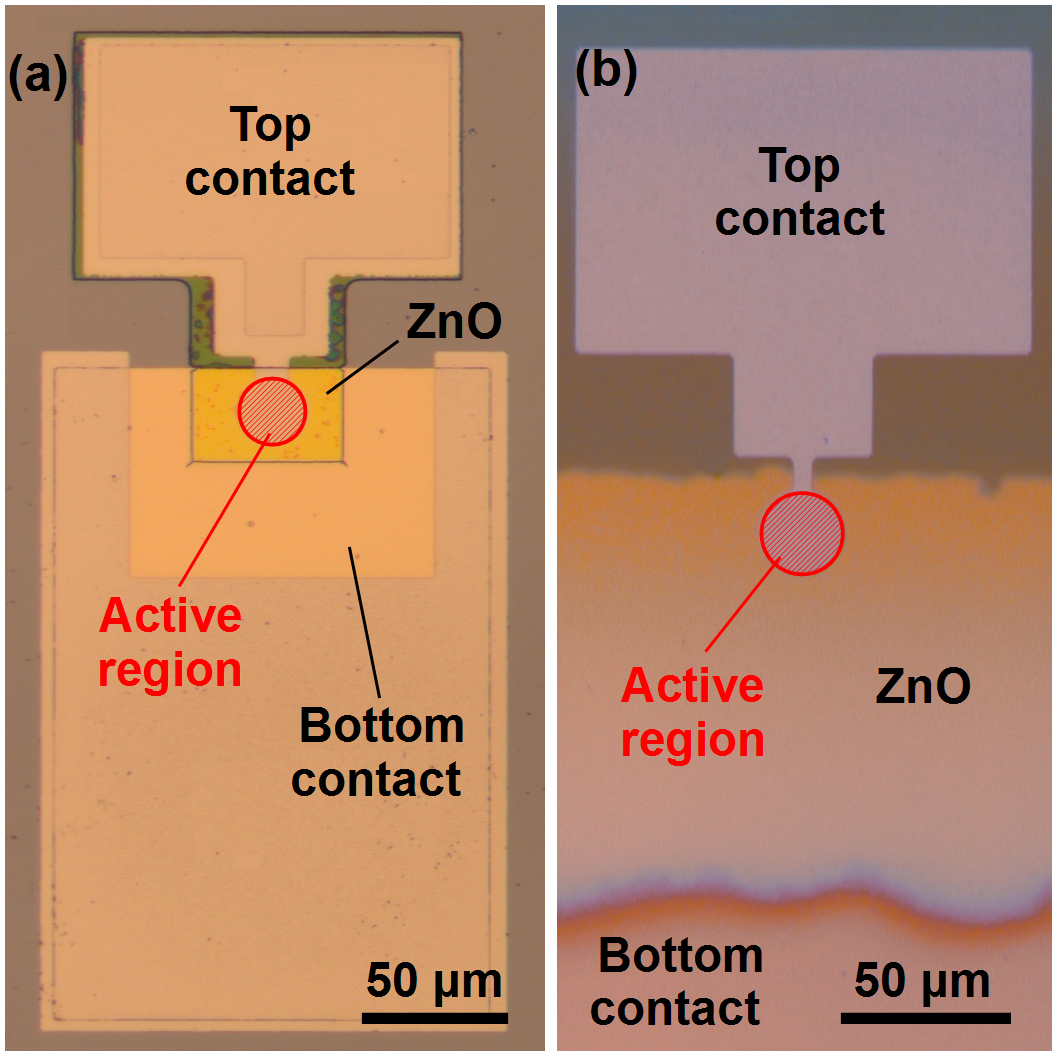}
    \caption{Optical micrographs of BAWRs fabricated using (a) optical lithography and (b) shadow-mask deposition. The active area of the BAWRs, where the top and bottom contacts sandwich the piezoelectric ZnO thin film, is indicated by a dashed red circle.}
    \label{Figure3}
\end{figure}

Two different approaches were used to define the lateral dimensions of the BAWR layers. The first was a conventional photolithographic process combining the lift-off deposition of the metallic contact layers with the wet-chemical etching of the sputtered ZnO film. In the second approach, the dimensions of the bottom contact as well as of the ZnO film were defined via evaporation and sputtering, respectively, through shadow masks. We used for that purpose mechanically machined shadow masks fabricated on a boron-nitride plate with minimum opening dimensions of approximately $100~\mu$m. The top contact, which defines the active area of the BAWRs, was fabricated by optical lithography.

Figure~\ref{Figure3} displays optical micrographs of the two types of structures indicating the contact areas (top and bottom contacts) as well as the piezoelectric ZnO film. In each case, the dashed red circle indicates the active area of the BAWRs. The main advantage of the shadow mask approach is the simplicity of fabrication, since it has less photolithographic steps and does not require the etching of the ZnO layers. Drawbacks of the shadow mask deposition method are the alignment tolerances ($\simeq 40~\mu$m) as well as the edge roughness of the layers [cf. Fig.~\ref{Figure3}(b)] resulting from the mechanical machining of the shadow masks, which are significantly larger than for photolithography. The photolithographic fabrication of the top contact together with a careful alignment to the underlying layers enable the fabrication of structures with active region dimensions down to approximately $10~\mu$m.

\subsection{Characterization techniques}

The structural properties of the BAWR layers were probed using atomic force microscopy (AFM). Their electrical response was investigated using a vector network analyzer to record the frequency dependence of the rf-scattering (s) parameters. The studies were carried out in a low-temperature probe station (temperatures from 10~K to room temperature, RT) with rf tips for contacting the BAWR devices. The setup has also calibration standards for the in-situ calibration of the scattering parameters.

\section{\label{Results}Results}

\subsection{\label{sec:Sec3_1}Structural properties of the ZnO films}

\begin{figure*}[t!bhp]
    \centering
    \includegraphics[width=1.90\columnwidth, angle=0, clip]{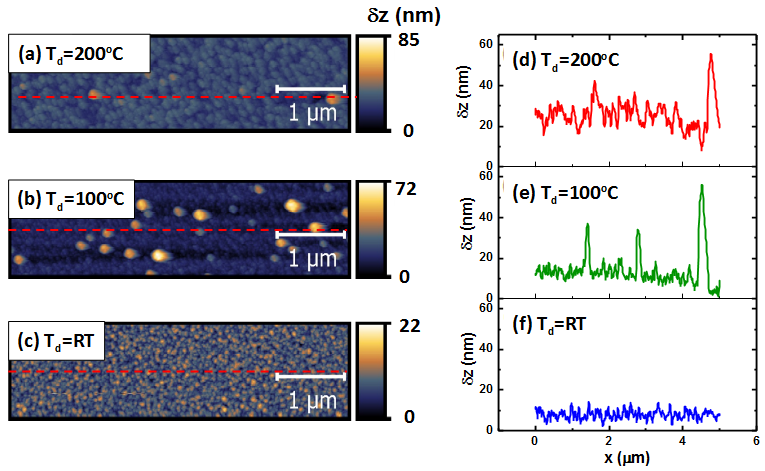}
    \caption{Atomic force micrographs (AFM) (left panels) and line profiles of the surface displacement  ($\delta z$, right panels) of BAWRs fabricated using 200~nm-thick piezoelectric ZnO layers sputtered at  temperatures (a)-(d) $T_d = 200^\circ$C,   (b)-(e) $T_d = 100^\circ$C, and (c)-(f) $T_d = \mathrm{RT}$. The line profiles were recorded along the dashed lines indicated \rPVS{in the left panels}{ on the left}.}
    \label{Figure4}
\end{figure*}

\begin{figure}[b!tpb]
    \centering
    \includegraphics[width=1.00\columnwidth, angle=0, clip]{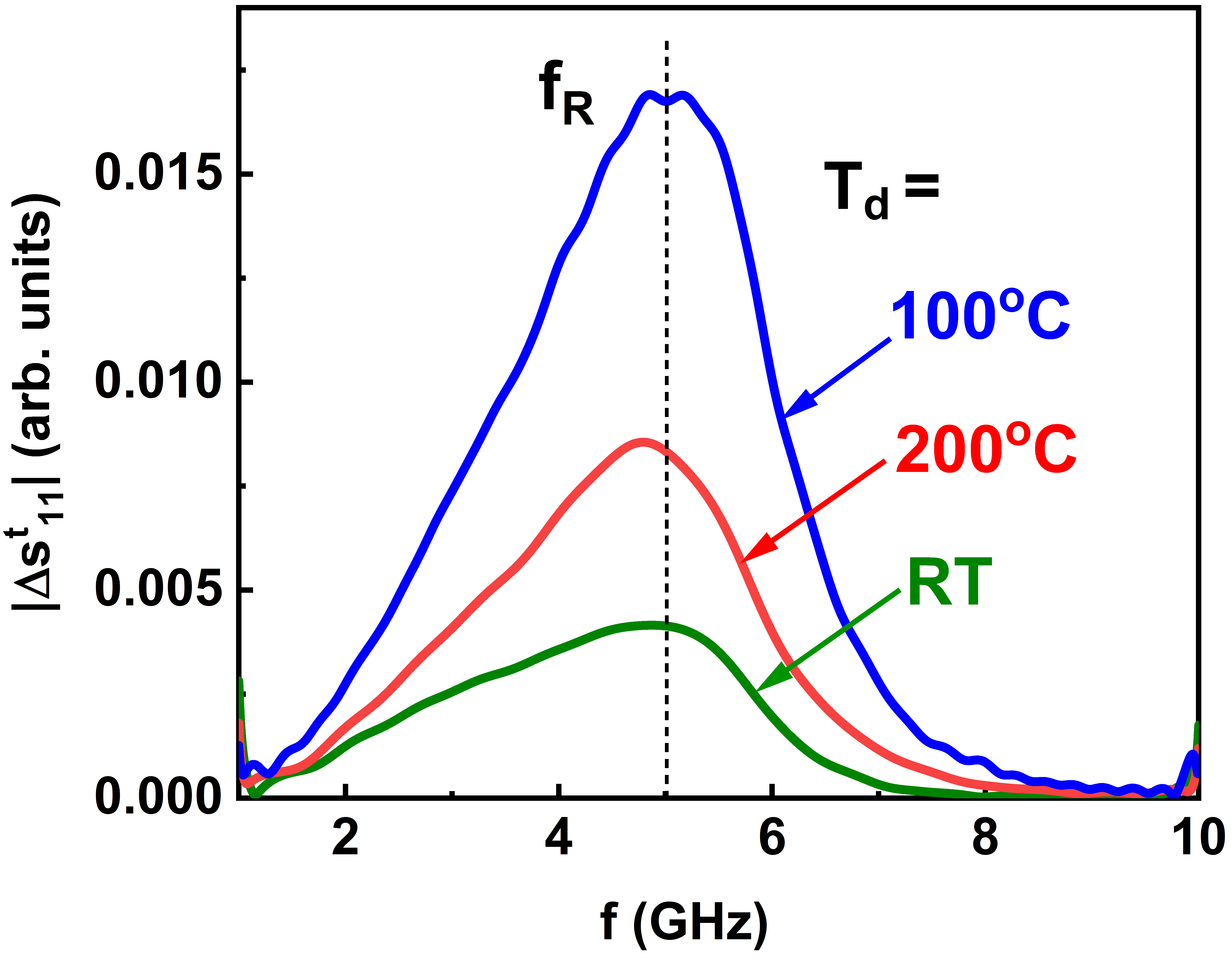}
    \caption{Dependence of the time-gated ${|\Delta s_{11,1}^t|}$ response of BAWRs with piezoelectric ZnO layers sputtered at different temperatures $T_d$. The BAWRs have an active area of $1960~\mu$m$^2$ and ZnO layers with a thickness of 440 nm. Here, $f_{R}$ indicates the resonance frequency.}
    \label{Figure5}
\end{figure}

The calculations of Sec.~\rPVS{\ref{sec:BAW_transducer_design}}{ 2.1} [cf. Fig.~\ref{Figure2}(b)] show that high-frequency ($>10$~GHz) BAWRs require ZnO layers with thicknesses below 100~nm. These layers should be smooth, highly resistive\aPVSn{, continuous} (i.e., without pinholes), and with good piezoelectric properties. 

 The structural and piezoelectric properties of the ZnO layers depend critically on the deposition conditions. The left panels in Fig.~\ref{Figure4} compare AFM micrographs maps of the vertical displacement $\delta z$ recorded on the surface of BAWRs with 200~nm-thick ZnO films sputtered at different temperatures $T_d$. The right panels display linear profiles of $\delta z$ recorded along the dashed lines indicated in left panels. 
Films deposited at RT [cf. Fig.~\ref{Figure4}(f)] have an average surface roughness  $r_a\sim 2$~nm (corresponding to the root-mean-square (rms) fluctuation of $\delta z$ \aPVS{averaged over the area of the micrograph}). For comparison, similar measurements on ZnO films deposited directly on the  GaAs (001) surface yield $r_a$ of only 0.5 nm. The same applies to the surface of the Au/Ti back contact used as the substrate for ZnO deposition, which has a roughness $r_a$ = 0.6 nm only marginally larger than the one of the GaAs substrates. The pronounced roughness of the BAWR surface is attributed to the columnar character of the textured ZnO layers, which is also a necessary condition for piezoelectricity.
The surface roughness increases with $T_d$ [cf. Figs.~\ref{Figure4}(d)-(e)]. In addition, the $\delta z$ profiles of  films deposited at high temperatures ($T_d\ge 100^\circ$C) exhibit sub-$\mu$ hillocks, which lead to spikes in $\delta z$ with amplitudes exceeding 20~nm. The mechanism for the appearance of these features was not further investigated.


 The deposition temperature of the ZnO films has a direct impact on the piezoelectric response of the BAWRs. Figure~\ref{Figure5} compares the time-gated rf-reflection coefficient \rPVSn{$\left|\Delta s_{11,1}^t\right|$}{~$\left| s_{11,1}^t\right|$} of BAWR structures designed for a resonance frequency $f_R = 5$~GHz with piezoelectric ZnO layers sputtered at different temperatures $T_d$. As will become clear in Sec.~\rPVS{\ref{Sec3_3}}{~3.3}, \rPVSn{$\left|\Delta s_{11,1}^t\right|^2$}{~$\left| s_{11,1}^t\right|$} is proportional to the electro-acoustic generation efficiency, thus being a figure of merit for the quality of the resonators. 
\rPVS{
The electro-acoustic performance considerably improves when the deposition temperature is increased from room temperature to  $100^\circ$C. This result is in agreement with previous studies of piezoelectric ZnO films for the excitation of SAWs on Si~\cite{PVS263} and GaAs~\cite{PVS309}, which show that the film properties increase with sputtering temperature (in \rPVSn{those}{~these} cases, the sputtering process was carried out at temperatures between 300 and 350$^\circ$C). A further increase of the sputtering temperature of the ZnO films on Au to $200^\circ$C results in rougher surfaces (cf. Fig.~\ref{Figure4}) as well as in a lower fabrication yield with a larger fraction of  short-circuited BAWRs. The latter indicates that the higher sputtering temperatures also induce the formation of pinholes.
 Based on these results, the studies reported in the subsequent sections were carried out using films sputtered at 100$^\circ$C, which yield the best electro-acoustic \rPVSn{performance}{~performances}.
}
{
The best electro-acoustic performance was obtained using films sputtered at $100^\circ$C.
 This temperature also yields the highest fabrication yield with the lowest density of shorted structures, thus indicating that the film roughness also induces the formation of pinholes. 
 Based on these results, the studies reported in the subsequent sections were carried out using films sputtered at 100$^\circ$C. 
}

\subsection{\label{Sec3_2}Radio frequency response of BAWRs}

Figure~\ref{Figure6}(a) displays the frequency dependence of the rf-reflected power  $\left|s_\mathrm{11}\right|^2$  (given in terms of the $s_\mathrm{11}$ rf scattering parameter) of a BAWR with a 440 nm-thick ZnO layer. The broad dip in the spectrum\rPVSn{}{~at} centered at approximately 5.55 GHz is attributed to the excitation of LBAWs: this resonance frequency corresponds closely to the one  calculated for the used ZnO thickness in Fig.~\ref{Figure2}(b). The spectrum also shows a series of oscillations (cf. figure inset), which are attributed to Fabry-Perot fringes induced by reflections of the LBAW at the back surface of the GaAs substrate \aPVS{(see details below)}. 
%
%
The period of the oscillations of 6.71 MHz agrees very well with  the expected value given by 
$ t_r^{-1}=v_\mathrm{GaAs}/(2d_\mathrm{GaAs})$, where $d_\mathrm{GaAs} = 356~\mu$m and $v_\mathrm{GaAs}= 4731$~m/s denote the thickness and longitudinal acoustic velocity in the GaAs substrate~\cite{LBIII_41a1a}, respectively.

The frequency response of the BAWR was modelled using the modified Butterworth-van-Dyke (BvD) equivalent electric circuit~\cite{Larson_IUS_863_00} depicted in Fig.~\ref{Figure6}(d). Here, the electro-acoustic transduction takes place in the right-most branch, where $R_a$ is the motional resistance quantifying the conversion of electric into acoustic power. The reactive components $C_a$ and $L_a$ are the motional capacitance and inductance, which are proportional to the inverse stiffness and density of the BAWR medium, respectively. The parallel branch includes the electrical capacitance across the electrodes $C_s=\epsilon_0\epsilon_\mathrm{ZnO}\frac{A_\mathrm{BAWR}}{d_\mathrm{ZnO}}$ in series with a resistance $R_s$, which accounts for dielectric losses. In the previous expression for $C_s$, $A_\mathrm{BAWR}$ is the effective area of the BAWR, and $\epsilon_0\epsilon_\mathrm{ZnO}$ the permittivity of the piezoelectric material. The BAWR components are connected  to an rf-source \rPVS{$V_\mathrm{rf}$}{$\mathrm{V_{rf}}$} with an internal series impedance $Z_0=50~\Omega$.

The lumped components $R_s$, $R_a$, $C_a$, and $L_a$ were determined by fitting the frequency response of the Butterworth-van-Dyke circuit to the measured $\left|s_\mathrm{11}\right|^2$ spectrum. The shunt capacitance $C_s$ was also fitted in order to account for small variations in thickness and dimensions of the BAWR. The solid line marked “BvD” in Fig.~\ref{Figure6}(a) displays the filter response calculated using the fitted values for the components listed in the caption.  The Butterworth-van-Dyke model reproduces reasonably well the measured resonance dip at the resonance frequency $f_R=\frac{1}{2\pi\sqrt{L_aC_a}}$. The dashed line “$R_s$-$C_s$“ shows, for comparison, the response of the $R_s$-$C_s$ parallel branch in the circuit of Fig.~\ref{Figure6}(d), which accounts for the dielectric losses of the BAWR.

The curve marked “BvD” in Fig.~\ref{Figure6}(c) displays the spectral dependence of the effective conversion efficiency $\Delta\left|s_\mathrm{11}\right|^2$ obtained from the difference between the dashed and solid curves in Fig.~\ref{Figure6}(a). The highest conversion value of  $\Delta\left|s_\mathrm{11}\right|^2\sim 0.35$ at the resonance frequency is mainly limited by the high acoustic impedance $R_a$, (approx. $250~\Omega$ for the present structure), which considerably exceeds the characteristic impedance $Z_0=50~\Omega$.

\subsection{\label{Sec3_3} BAWR transduction efficiency}

The analysis based on the Butterworth-van-Dyke model of the previous sections yields information about the overall electro-acoustic conversion efficiency. In this section, we used rf-measurements in the time-domain to estimate the fraction of the acoustic power transduced to propagating LBAWs.

\begin{figure*}[t!]
    \centering
    \includegraphics[width=1.90\columnwidth, angle=0, clip]{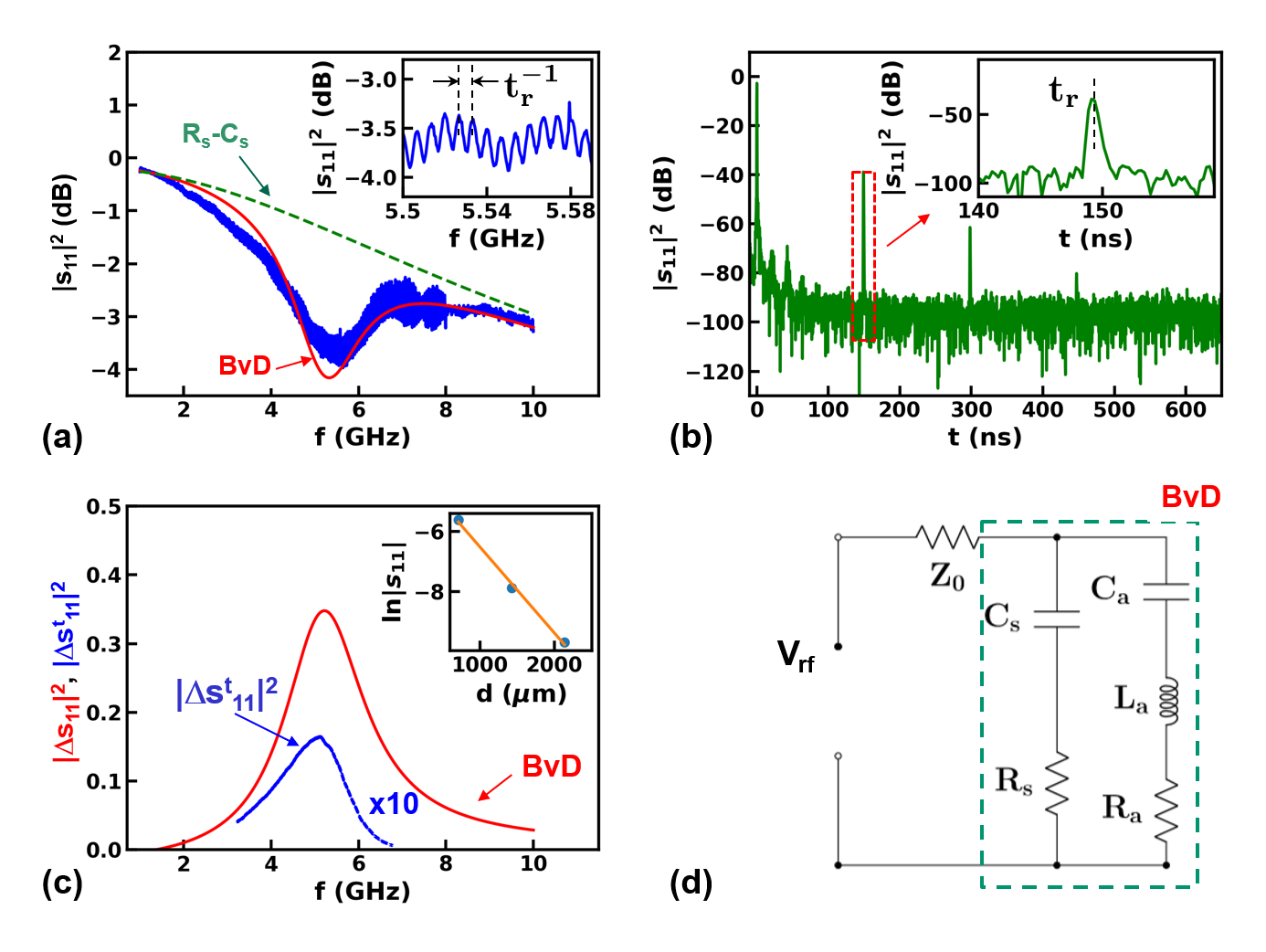}
    \caption{(a) Frequency dependence of  $\left|s_{11}\right|^2$  at room temperature for a BAWR with a 440 nm thick ZnO film. The solid line is a fit to the Butterworth-van-Dyke (BvD) circuit model shown in (d), while the dashed line represents the response due to the \rPVS{$C_{s}$-$R_s$}{~$C_s-R_s$} shunt branch.  The magnified region in the inset shows Fabry-Perot-like oscillations with frequency $t_r^{-1}$. (b) Time domain traces for $\left|s_{11}\right|^2$ displaying the echoes due to LBAW reflection at the backside of the substrate. The inset displays details of the first echo at time $t_r$. (c) Effective electro-acoustic conversion efficiency $|\Delta s_\mathrm{11}|^2$ determined from the difference between the curves marked “BvD” and “Rs-Cs” in (a). The line marked ``$|\Delta s^t_\mathrm{11}|^2$'' is the rf-reflectivity contribution from LBAW modes determined from the intensity of the echoes in (b), which are plotted as a function of the echo propagation distance $d$ in the inset. (d) Butterworth-van-Dyke lumped filter model used to fit the rf-spectrum in (a), yielding the following parameters: $R_a= 260~\Omega$, $R_s=18~\Omega$, $C_a=4.4$~fF, and $L_a=22$~nH.  
}
    \label{Figure6}
\end{figure*}

The time-domain trace for $\left|s_\mathrm{11}\right|^2$ of Fig.~\ref{Figure6}(b) shows a series of acoustic echoes, which are associated with the Fabry-Perot oscillations depicted in the \rPVS{inset}{~caption} of Fig.~\ref{Figure6}(\rPVSn{a}{b}). This trace was determined by Fourier transforming the frequency response in the frequency range from 3 to 7~GHz. Three echoes at times $i\times t_r\left(i=1,2,3\right)$ corresponding to multiple round trips across the substrate with the LBAW velocity can be observed in this case. 
The inset of Fig.~\ref{Figure6}(c) shows that the integrated amplitude of the \aPVS{$|s_\mathrm{11}|^2$} echoes decays exponentially according to \rPVS{ $r_\mathrm{eff}=\exp{(-2 \alpha_\mathrm{eff}d_\mathrm{GaAs})}$}{~$e^{-2 \imath \alpha_\mathrm{eff}d_\mathrm{GaAs}}$}. Here,  $d_\mathrm{GaAs}$ is the substrate thickness and $\alpha_\mathrm{eff}$  \rPVS{the effective acoustic absorption coefficient, defined as the inverse of the LSAW propagation distance for an amplitude decay by a factor $e^{-1}$ ($e$ is Euler's number). Note that  $\alpha_\mathrm{eff}$ is an effective coefficient since it includes not only propagation losses but also losses due to scattering at surfaces and interfaces. }{effective amplitude decay length of the LBAW.}
 From the plot, we extract   $\alpha_\mathrm{eff}=0.0029~\mu$m$^{-1}$ as well as the exponential \aPVS{amplitude} decay factor per round trip across the substrate 
\rPVS{$r_\mathrm{eff} = 0.13$}
{~$r_\mathrm{eff}= 2\alpha_\mathrm{eff}\times d_\mathrm{GaAs}=0.13$} for 5~GHz LBAWs propagating at RT. As will become clear in the subsequent sections, the progressive amplitude decay of the echoes at RT is mainly due to acoustic losses during propagation through the substrate.


The effective electric response $\Delta s^t_\mathrm{11}$ arising from the electrical reabsorption of the LBAW modes by the BAWR can be obtained by summing the time-gated contributions from the echoes according to:  
\begin{eqnarray}
\label{Eq1}
\Delta s_\mathrm{11}^t &=& \sum_{i=0}^{ \infty }{ s_\mathrm{11,i}^t  \left(  e^{ 2 \pi  \rPVSn{j}{\imath} {f}{t_r}} r_\mathrm{eff} \right)^i} 
   =  s_\mathrm{11,0}^t \frac{1}{1 - e^{ 2\pi  \rPVSn{j}{\imath} {f}{t_r} } r_\mathrm{eff}}\\ \nonumber 
&=& 
s_\mathrm{11,1}^t \frac{1}{ r_\mathrm{eff} \left[ 1 - e^{ 2\pi \rPVSn{j}{\imath}   {f}{t_r} } r_\mathrm{eff}\right]}, 
\text{with } \rPVSn{j}{\imath}=\sqrt{-1}.
\end{eqnarray}

\noindent \rPVS{In this expression, the  exponential factor containing the LBAW frequency ($f$) yields the phase shift accumulated by the wave during a round trip across the substrate. The closed forms on the right-hand-side were obtained by using the expression for the summation of a convergent infinit\rPVSn{e}{y} series.}{~In this expression.} \aPVS{For $i > 0$,} $s_\mathrm{11,i}^t$ denotes the component of the frequency response associated with the $i^{th}$-echo in the time-domain traces for $s_\mathrm{11}$.
%
\rPVS{These components }{For $i > 0$, $s_\mathrm{11,i}^t$}  can be directly obtained from time-domain traces: this procedure is normally used to discriminate the acoustic response from the much faster electromagnetic one. The total response ${\Delta s}_\mathrm{11}^t$ also includes the response $s_\mathrm{11,0}^t$ at short times (i.e., before the LBAW leaves the transducer region), which overlaps with the much stronger electromagnetic reflection at zero times, thus normally remaining inaccessible. The last term at the right-hand-side of Eq.~\ref{Eq1} shows that the total rf-response associated with the LBAW can be retrieved by correcting $s_\mathrm{11,1}^t$ by a factor that depends on  $r_\mathrm{eff}$.

${\Delta s}_\mathrm{11}^t$ represents the frequency-dependent fraction of the input electrical power, which is transduced to a propagating LBAW mode and subsequently reconverted to an electric signal. 
The latter requires that the LBAW propagates without wavefront distortions, which can arise from diffraction or reflections on irregular surfaces, since these will reduce the reconversion efficiency to an electrical signal by the BAWR. 
${\Delta s}_\mathrm{11}^t$ is thus a direct measure of the BAWR transduction efficiency as well as of the ability of the LBAW mode to propagate without absorption and wavefront distortions.

\aPVS{
If $r_\mathrm{eff}$ in Eq.~(\ref{Eq1}) is small, then the amplitude of the Fabry-Perot oscillations in Fig.~\ref{Figure6}(a) can be approximated by 
$|\Delta s_\mathrm{11}^t|^2  \sim  |s_\mathrm{11,0}^t|^2  \left[ 1  +  2r_\mathrm{eff}\cos{ \left(2\pi {f}{t_r}\right)} \right]$. This expression shows that the oscillations are sinusoidal in this approximation with an amplitude proportional to $2r_\mathrm{eff}$.}
Equation~(\ref{Eq1}) \aPVS{also} provides a convenient route to determine the LBAW generation efficiency by high-frequency ($>5$~GHz) ZnO/GaAs BAWRs, where the low electro-acoustic conversion efficiency leads to small dips in the ${{|s}_\mathrm{11}|}^2$ spectra. Note that, in general, one needs to take into account the frequency dependence of $r_\mathrm{eff}$. The latter can, however, be neglected if the time traces are calculated for a small frequency range $\Delta f_R$ of the $s_\mathrm{11}$ spectrum around the BAWR resonance.

The line marked $\left|\Delta s_\mathrm{11}^t\right|^2$ in Fig.~\ref{Figure6}(c) displays the rf-reflectivity $\left|{\Delta s}_\mathrm{11}^t\right|^2$ calculated from Eq.~(\ref{Eq1}) using the frequency independent $r_\mathrm{eff} = 0.13$ determined from the amplitude of the echoes (cf. inset). The peak position as well as line shape of the $\left|{\Delta s}_\mathrm{11}^t\right|^2$ resonance resembles the one determined from the fits to the Butterworth-van-Dyke circuit model (solid line). The peak intensity of approx. 1.5\%  is, however, much lower than the 35\% peak value of $\left|\Delta s_\mathrm{11}\right|^2$ value obtained by the Butterworth-van-Dyke analysis. We thus conclude that only a small fraction of the electro-acoustical power of the propagating LBAW mode is recovered by the transducer.

According to the calculations shown in Fig.~\ref{Figure2}(b), the resonance frequency $f_R$ of the BAWRs can be increased by reducing the thickness of the piezoelectric layer. This behavior is experimentally confirmed by the dots in this figure, which display the center frequency of the $|s_\mathrm{11,i}^t|^2$ resonances measured at RT. While the calculations reproduce very well the resonance frequencies for large $d_\mathrm{ZnO}$, the experimental points lie above the calculations for small ZnO thicknesses. We attribute this discrepancy to  slightly deviations between the  real and nominal thicknesses of the Au/Ti bottom contact in these structures.

\subsection{\label{Sec3_4}Temperature dependence of the LBAWR transduction}


\begin{figure*}[t!]
    \centering
    \includegraphics[width=1.90\columnwidth, angle=0, clip]{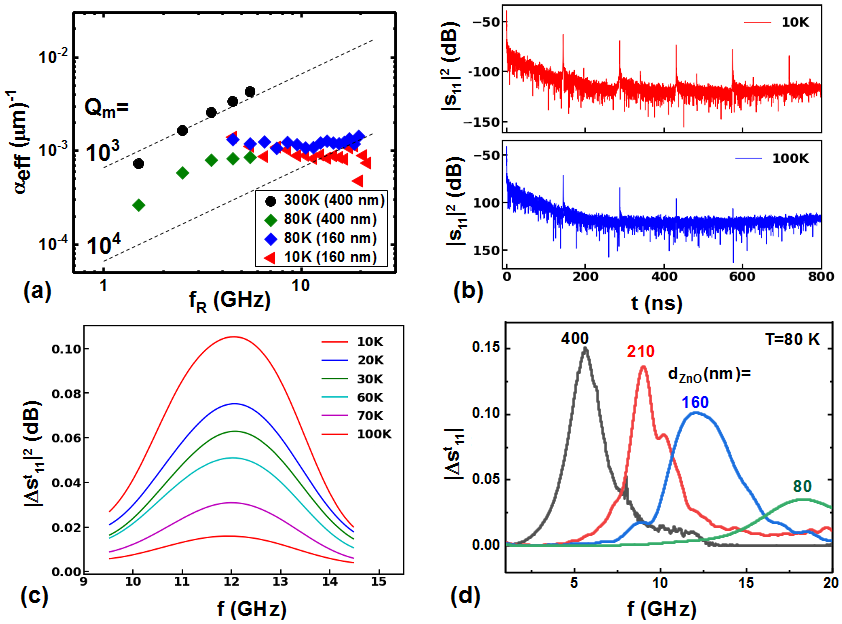}
\caption{(a) Frequency dependence of the LBAW absorption coefficient $\alpha_\mathrm{eff}$ at different temperatures. The dashed line shows the frequency dependence of $\alpha_\mathrm{eff}$ required to yield the indicated quality factors $Q_M$ defined in Eq.~\ref{Eq2}.
(b) Impulse response of a BAWR with a $(135\pm5)$~nm thick ZnO layer (active area of $490~\mu$m$^2$) recorded at 10~K and 100~K. The time-domain traces were determined from \rPVS{$s_\mathrm{11}$}{$s\mathrm{11}$} spectra recorded over a 6 GHz band around the resonance frequency $f_R =12$~GHz. 
(c) Temperature dependence of the $|\Delta s^t_\mathrm{11}|^2$ response for a BAWR with $f_R=12$~GHz \aPVSn{(the temperatures are listed in the legend)}. 
(d)  $|\Delta s^t_\mathrm{11}|^2$ spectra recorded at 80K for BAWRs fabricated with different ZnO layer thicknesses $d_\mathrm{ZnO}$. 
}
    \label{Figure7}
\end{figure*}


The dots in Fig.~\ref{Figure7}(a) display the dependence of the effective absorption coefficient $\alpha_\mathrm{eff}$ on frequency, as determined at RT by analyzing the amplitude of the $\left|s_\mathrm{11}\right|^2$ echoes for frequency slices $\Delta f_R = 1$~GHz. The increase of the product $d_\mathrm{ZnO}\alpha_\mathrm{eff}$ with frequency limits the frequency range for the observation of echoes to approx. 10~GHz (note that the determination of $\alpha_\mathrm{eff}$ requires at least two echoes in the time domain traces). This frequency range can be considerably extended by performing the measurements at low temperatures. Figure~\ref{Figure7}(b) compares time domain traces for $s_\mathrm{11}$ recorded on a structure for $f_R = 12$~GHz at 10~K and 100~K. Note that the intensity and the number of \aPVSn{detected} echoes increase\rPVSn{}{s} substantially in the former case, thus indicating a much \rPVSn{lower}{~weaker} acoustic absorption coefficient $\alpha_\mathrm{eff}$ at low temperatures. The reduction of  $\alpha_\mathrm{eff}$ is accompanied by a large enhancement of $|\Delta s_\mathrm{11}^t|^2$, as illustrated in Fig.~\ref{Figure7}(c) for a BAWR with $f_R = 12$~GHz. \rPVSn{It is evident}{~Note} that there is an almost ten-fold increase of  $|\Delta s_\mathrm{11}^t|^2$ as the temperature is decreased from 100~K to 10~K. At low temperatures, it thus becomes possible to detect multiple echoes for LBAWs with frequencies exceeding 20~GHz. As an example, Fig.~\ref{Figure7}(d) compares  $|\Delta s_\mathrm{11}^t|^2$ spectra of BAWRs with ZnO layers thickness varying from 80  to 400~nm. The structure with the thinner ZnO layer resonates at 18.4~GHz with a bandwidth of approx. 5~GHz. To our knowledge, the latter is the highest reported resonance frequency for BAWR on GaAs.

\subsection{\label{Sec3_5}BAWRs on distributed Bragg reflectors}

\begin{figure*}[htpb]
    \centering
    \includegraphics[width=1.90\columnwidth, angle=0, clip]{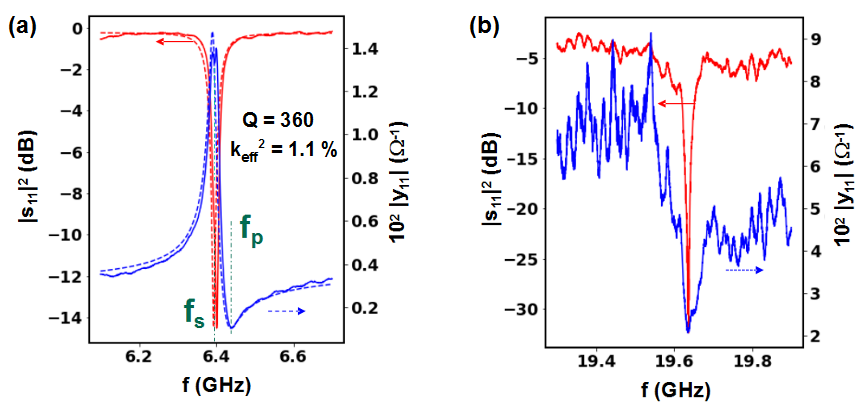}
\caption{Electric response at 10 K of BAWRs with resonant frequencies of (a) 6.4 GHz and (b) 19.64 GHz deposited on (Al,Ga)As distributed Bragg reflectors (DBRs). The solid (red) and dashed (blue) curves display the measured and fitted values to the $|s_\mathrm{11}|^2$ (left scale) and $|y_\mathrm{11}|$ spectra, respectively. The fits were carried out to the  Butterworth-van-Dyke model of \rPVS{Fig.~6(d)}{8(d)}. The nominal ZnO thicknesses in (a) and (b) are 440 nm and 70 nm, respectively. $f_s$ and $f_p$ in (a) denote the serial and parallel resonance frequencies of the structures used to determine the quality factor Q and the effective coupling coefficient $k^2_\mathrm{eff}$. 
}
    \label{Figure8}
\end{figure*}

 The quality factor of BAWRs becomes significantly enhanced when they are deposited on distributed Bragg reflectors (DBR) to prevent wave leakage to the substrate~\aPVS{\cite{Newell_PIEEE53_575_65,Lakin_2_2_95,Trigo02a}}. Figures~\ref{Figure8}(a) and \ref{Figure8}(b) compare spectra recorded at 10~K of the scattering ($|s_\mathrm{11}|^2$) and admittance ($|y_\mathrm{11}|$) parameters of BAWRs with resonant frequencies $f_R$ of 6.4 GHz and 19.64 GHz, respectively. These BAWRs were deposited on DBRs consisting of Al$_{x_1}$Ga$_{(1-x_1)}$As/Al$_{x_2}$Ga$_{(1-x_2)}$As layer stacks with different Al contents $x_i$, each with a thickness corresponding to $1/4$ of the resonance acoustic wavelength. The DBRs were grown on (100) GaAs substrates using molecular beam epitaxy. For the $f_R = 6.4$~GHz structure, the DBRs consist of 10 stack pairs with Al contents $x_1 = 0$ and $x_2 = 0.85$\aPVS{ and thicknesses of 179~nm and 210~nm, respectively}. For the second structure, the DBR has 20 layer pairs with compositions \rPVS{$x_1 = 0.05$ and $x_2 = 0.85$ and corresponding thicknesses of 59.7~nm and 70~nm, respectively.}{~$x_1 = 0.05$\% and $x_2 = 0.85$\%.}

For both BAWR in Fig.~\ref{Figure8}, the $|s_\mathrm{11}|^2$ spectrum (solid curve) shows a  minimum at $f_R$ with a much narrower half width than in Fig.~\ref{Figure7}(d). The  difference arises from the fact that DBRs confine the vibrations in a region close to the BAWRs. As a result, vibrations produced at  successive rf-cycles interfere constructively, thus leading to a higher amplitude.

In order to determine the acoustic quality factor $Q$ and the effective electro-mechanical coupling coefficient $k_{\mathrm{eff}}^2$, we also show in Fig.~\ref{Figure8} the admitance  ($|y_\mathrm{11}|$) spectra of the structures~\cite{Hashimoto_Book_09}. The latter display a maximum followed by a minimum corresponding to the serial and parallel resonance frequencies $f_s$ and $f_p$, respectively\dPVS{~\cite{Hashimoto_Book_09}}. The dashed lines in Fig.~\ref{Figure8}(a) show the values for $|s_\mathrm{11}|$ and $|y_\mathrm{11}|$ obtained by fitting the experimental data to the modified Butterworth-van-Dyke model of Fig.~\ref{Figure6}(d). From the values for the lumped circuit components obtained from the fits we determined \rPVS{the resonance frequencies $ f_p=  \left(2 \pi \sqrt { L_a (C_a+C_s)/(C_a C_s)} \right)^{-1}$ 
and $f_s=f_R = \left(2 \pi \sqrt{L_a C_a}\right)^{-1}$~\cite{Tirado_PhD_10,Hashimoto_Book_09},   
the quality factor}{ the quality factor }:
\begin{equation} 
Q=\frac{1}{2} \frac{f_p+f_s}{f_p-f_s},
\end{equation}

\noindent as well as the effective coupling coefficient:
\begin{equation} 
k_{\mathrm{eff}}^2=\frac{\pi}{2} \frac{f_s}{f_p} \tan{\left( \frac{\pi}{2} \frac{f_s}{f_p} \right)}
\end{equation}

\noindent for the sample in (a) following the procedure \rPVSn{described}{~delineated} in Ref.~\onlinecite{Hashimoto_Book_09}. We obtained $k_{\mathrm{eff}}^2=1.1$\% as well as a quality factor $Q=360$. The latter is most likely determined by leakage of the LBAW \rPVSn{into}{~to} the substrate due to the finite thickness of the DBR rather than by acoustic absorption within the DBR region. 

\rPVS
{The 19.6~GHz sample in Fig.~\ref{Figure8}(b) has a narrower $|s_\mathrm{11}|^2$ resonance, which is probably due to the larger number of DBR pairs and, therefore, reduced wave leakage to the substrate. Its  $|s_\mathrm{11}|^2$ response has, however, a  higher baseline ($\approx -5$~dB)  than the one for the 6.4~GHz sample. In addition,  the fitting procedure used before could not be reliably applied to extract the Q-factor and the electromechanical coupling $k_{\mathrm{eff}}^2$. The reasons for this behavior are currently under investigation. 
 }
{Due to the \aPVS{weaker} $|y_\mathrm{11}|$ response, a similar fitting procedure could not be reliably carried out for the 19.6~GHz sample in \ref{Figure8}(b). The narrower width of the $|s_\mathrm{11}|$ resonance in this case is attributed to the larger number of DBR pairs, which reduces wave leakage to the substrate. }

\section{\label{Discussions}LBAW absorption in GaAs}

The results of the previous sections set the base for the piezoelectric generation of SHF LBAWs in GaAs. In this section, we turn our attention to the mechanisms determining the damping of LBAWs in GaAs in the SHF-range as well as to the consequences for high quality acoustic resonators operating in this range. 

\subsection{\label{Sec4_1}Quality factor and acoustic absorption mechanisms}

The temperature and frequency dependences of $\alpha_\mathrm{eff}$ are summarized by the symbols in Fig.~\ref{Figure7}(a). At RT, the effective absorption follows a power-law dependence $\alpha_\mathrm{eff} \sim f^m$ with an exponent $m$ between 1 and 2. At low temperatures, in contrast, $\alpha_\mathrm{eff}$ becomes essentially independent of frequency within the investigated frequency range, thus yielding an exponent $m \sim 0$.

Assuming that the LBAWs propagate with the GaAs longitudinal velocity $v_{\mathrm{GaAs}}$, \dPVS{the inverse amplitude decay length~}$\alpha_\mathrm{eff}$ becomes related to the BAW decay time $\tau_p$ according to:
\begin{equation}
\tau_p=\frac{1}{ 2\alpha_\mathrm{eff}v_\mathrm{GaAs}}. 
\label{Eqtau}
\end{equation}

\noindent The factor of two in the denominator accounts for the fact that $\tau_p$ is defined as an energy (rather than amplitude) loss rate. Using Eq.~(\ref{Eqtau}) one obtains the following expression for the upper limit $Q_M$ for the quality factor of a GaAs acoustic resonator operating at frequency $f_R$~\cite{Daly_PRB80_174112_09}:  
\begin{equation}
Q_M=\frac{f_R}{\Delta f_R}= {2\pi f_R}{\tau_p}
=   \frac{\pi}{v_\mathrm{GaAs}}\frac{f_R} { \alpha_{\mathrm{eff}}}.
\label{Eq2}
\end{equation}

The dashed lines in Fig.~\ref{Figure7}(a) show the expected frequency dependence of $\alpha_\mathrm{eff}$ required for achieving the quality factors $Q_M$ indicated in the figure. Due to the exponent $m>1$ in the frequency dependence of $\alpha_\mathrm{eff}$ at RT, $Q_M$ is expected to decrease with increasing $f_R$. For low temperatures, in contrast, $m\sim 0$ and one reaches the favorable situation in which the contribution to quality factor associated with acoustic absorption reduces with  increasing $f_R$.

\begin{figure*}[htbp]
    \centering
    \includegraphics[width=1.90\columnwidth, angle=0, clip]{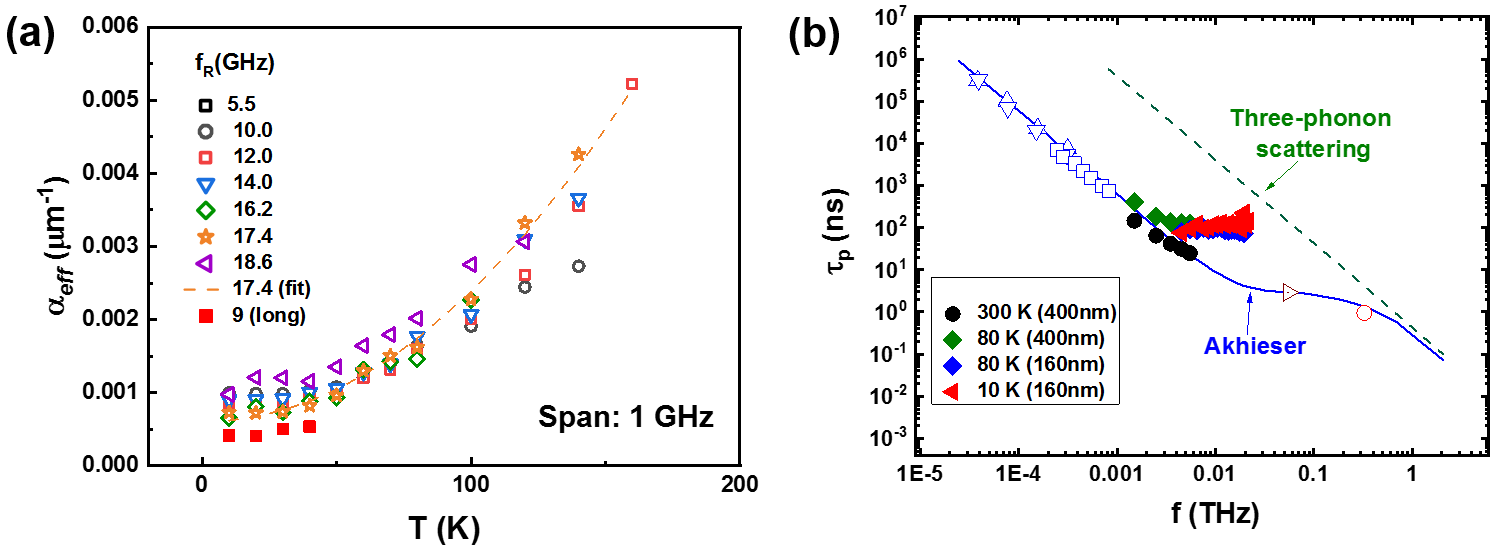}
\caption{
 (a) Temperature dependence of  $\alpha_\mathrm{eff}$ measured in BAWRs with different resonance frequencies $f_R$. The dashed line is a parabolic fit to the data for $f_R = 17.4$~GHz (cf. Eq.~(\ref{Eq2})).
 (b) Frequency dependence of the LBAW lifetime $\tau_p$. The solid symbols show the results from the present studies for different temperatures. 
 The open symbols for frequencies below 1~GHz display measured results obtained by ultrasound propagation studies along the GaAs $\langle100\rangle$ (downward triangles, Ref.~\onlinecite{Cottam_JPCSSP7_2447_74}) 
 and  $\langle111\rangle$ directions (upward triangles, Ref.~\onlinecite{Cottam_JPCSSP7_2447_74} and open squares, Ref.~\onlinecite{Ivanov_SPSS13_508_71}). The open circle and right facing triangle are laser picosecond results from Refs.~\onlinecite{Chen_PMB70_687_94} and \onlinecite{Maznev_APL102_41901_13}. The solid and dashed lines are the predictions from the Akhieser and three-phonon model from Ref.~\onlinecite{Maznev_APL102_41901_13}, respectively.
}
    \label{Figure9}
\end{figure*}

To address the mechanisms responsible for acoustic absorption $\alpha_\mathrm{eff}$ for propagating LBAWs in the GaAs, we summarize in Fig.~\ref{Figure9}(a) the measured temperature dependence of $\alpha_\mathrm{eff}$  for LBAW with different frequencies. In all case, $\alpha_\mathrm{eff}$  exhibits a flat region at low temperatures followed by a second region with an approximately quadratic dependence on temperature. This behavior can be summarized by the following expression:
\begin{equation}
\label{Eq3}
\alpha_\mathrm{eff}\left(f,T\right)=\alpha_\mathrm{eff,0}\left(f\right)+\ \alpha_\mathrm{eff,1}(f) \left[\frac{T}{300\ K}\right]^2. 
\end{equation}

\noindent Here, $\alpha_\mathrm{eff,0}$ is the saturation value of the absorption at low temperatures while $\alpha_\mathrm{eff,1}$ denotes the coefficient of the temperature-dependent term $T^2$. 
The dashed line in Fig.~\ref{Figure9}(a) is a fit of Eq.~(\rPVS{\ref{Eq3}}{\ref{Eq2}}) to the data for the $f_R = 17.4$~GHz structure yielding $\alpha_\mathrm{eff,0} = 6\times 10^{-4}~\mu m^{-1}$ and $\alpha_\mathrm{eff,1} = 0.016~\mu m^{-1}$.

The effective coefficient $\alpha_\mathrm{eff} (f)$ can also be expressed as a sum of contributions from different loss mechanisms according to:
\begin{eqnarray}
\label{Eqalphaeff}
\alpha_\mathrm{eff}\cong &\alpha_\mathrm{prop} +\\ \nonumber
 &\frac{1}{d_\mathrm{GaAs}} \left[ \ln{r_{TS}} + \ln {r_{BS}} + \ln{ r_e} + \ln{r_{misc}} \right].
\end{eqnarray}

To simplify the notation, we have omitted the functional dependence of the different terms of Eq.~(\ref{Eqalphaeff}) on frequency and temperature. In this equation,  $\alpha_\mathrm{prop}$ is the intrinsic absorption losses of the LBAW during propagation through both the BAWR and GaAs substrate. The terms within the square brackets account for losses due to surfaces, interfaces and beam diffraction. $r_\mathrm{BS}$ denotes the losses due to reflection at the bottom surface as well as leakage to the underlying sample holder. The latter depend on the surface roughness as well as on the acoustic coupling to the underlying medium. The term $r_\mathrm{TS}$ quantifies the effective losses during the propagation through the BAWR as well as scattering at its surface and interfaces while $r_e$ accounts for the acoustic losses per round trip due to the acousto-electric conversion within the BAWR. The last term $r_\mathrm{misc}$ encompasses additional loss mechanisms, such as diffraction, non-parallelism of the wafer surfaces, etc.

\subsection{\label{Sec4_2}Acoustic losses at room temperature}

Except for  the first contribution $\alpha_\mathrm{prop}$, all terms on the right-hand-side of  Eq.~(\ref{Eqalphaeff}) are expected to be weakly temperature dependent, thus indicating that this term dominates  at RT. Furthermore, since the wave propagation distance within the BAWRs is over three orders of magnitude shorter than in the GaAs substrate, we will assume that propagation losses within the BAWR are smaller than in the substrate (for this reason, we used in Eq.~(\ref{Eqalphaeff}) the substrate thickness $d_\mathrm{GaAs}$ rather than the total sample thickness). Consequently, the temperature-dependent contribution in Eq.~(\ref{Eq3}), which dominates the losses at RT, must thus be related to the intrinsic LBAW absorption during propagation through the GaAs substrate. In fact, several LBAW propagation studies in GaAs have reported a quadratic dependence of $\alpha_\mathrm{eff}$ like the one in Eq.~(\ref{Eq3}). The measured coefficients $\alpha_\mathrm{eff,1}(f)$ range from $0.04~\mu$m$^{-1}$ for 56 GHz LBAWs~\cite{Chen_PMB70_687_94} to $0.25~\mu$m$^{-1}$ for 713~GHz waves~\cite{Markov_PRB98_245201_18}. The value $\alpha_\mathrm{eff,1}(f) = 0.016~\mu$m$^{-1}$ determined in Sec.~\ref{Sec4_1} for frequencies between 5 and 20~GHz follows well this trend with frequency. Such a dependence has been associated with the interaction of the LBAW with thermal phonons. 

Figure~\ref{Figure9}(b) compares the LBAW lifetime $\tau_p=1/\left(2\alpha_\mathrm{eff}v_\mathrm{GaAs}\right)$ (as previously mentioned, the factor 2 arises from the fact that $\tau_p$ is defined as an intensity decay time) obtained in the present studies at different temperatures (solid symbols) with previously reported data (adapted from Fig. 4 of Ref.~\cite{Maznev_APL102_41901_13}\aPVSn{)}. 
The data below 1~GHz was determined by ultrasound absorption while points above 50~GHz \aPVSn{were} obtained using laser acoustics. 
The SHF LBAW propagation studies reported here cover the frequency window for the acoustic absorption of LBAW modes from 1 to 20~GHz, which has so far not been experimentally accessed in GaAs. 
The solid and dashed lines in Fig.~\ref{Figure9}(b) are the predictions of the Akhieser and three-phonon models, respectively~\cite{Maznev_APL102_41901_13}. These models describe the mechanism of LBAW damping associated with the  interaction with high-frequency thermal phonons.
Note that our present data set for RT (filled circles) is fully consistent with previous experimental results as well as with the predictions of the Akhieser model. According to this model, the strain field of the LBAW changes the energy distribution of thermal phonons via the Gr\"{u}neisen parameters. The thermal vibrations readjust their energy distribution via anharmonic effects on a time scale short in comparison with the LBAW period, thus leading to an irreversible energy transfer to the thermal bath~\cite{Maris_physacoustics8_279_71,Duquesne_PRB68_134205_03}.

\subsection{\label{Sec4_3}Acoustic losses at low temperatures}

At low temperatures, the acoustic absorption reduces significantly and saturates at a frequency dependent value $\alpha_\mathrm{eff,0}$ (cf. Fig.~\ref{Figure9}). The latter is expected to be determined by the second term on the right-hand-side of Eq.~(\ref{Eqalphaeff}). In order to identify the mechanisms responsible for acoustic losses at low temperatures, we performed two additional control experiments. The first one addressed losses associated with  reflections at the back surface ($r_\mathrm{BS}$) as well as with leakage to the underlying sample holder. For that purpose, we measured $\alpha_\mathrm{eff}$ at RT for a BAWR with $f_R = 5$~GHz under different mounting conditions: (i) suspended, (ii) in contact with the substrate holder, and (iii) glued to the substrate holder. The changes in  ~$\alpha_\mathrm{eff}$ were of less than 10\% and, thus, smaller than $\alpha_\mathrm{eff,0}$ in Fig.~\ref{Figure7}(a). This result indicates that this  mechanism has a minor role in the propagation losses at RT. The experimental studies, however, \rPVSn{could}{~can} not be extended to low temperatures.

In a second set of experiments, we have carried out acoustic absorption measurements at low temperatures over much longer propagation distances $d_\mathrm{GaAs}$. As can be inferred from Eq.~(\ref{Eqalphaeff}), a larger $d_\mathrm{GaAs}$ increases the relative weight of the intrinsic absorption (first term on the \rPVS{right-hand-side}{~rhs}) over the other terms, thus indirectly yielding information about the role of surface and interfaces. 
\rPVS{The experiments were carried out by exciting LBAWs using BAWRs fabricated on a (110) GaAs slab. The substrates were from the same batch as the ones used for the previous studies, but  the LBAWs now propagate along the $\langle 110 \rangle$ direction.     In addition, while the upper and bottom surfaces of the (100) GaAs wafers were polished, those of the (110) GaAs slabs were cleaved, which may result in different acoustic reflection properties. }
{The experiments were carried out by exciting BAWs on a 1.5 mm-thick (110) GaAs substrate with cleaved upper and bottom surfaces. The GaAs substrates employed in these experiments were similar to the ones used for the $\langle 100 \rangle$ propagation studies. }

\rPVS{The solid squares in Fig.~\ref{Figure9}(a) show the effective absorption coefficient measured on  $1.5~\mu$m-thick $\langle110\rangle$ GaAs slab.  Due to the thicker substrates, the
}{The}  echo propagation lengths \aPVS{in these samples} are \dPVS{thus} approx. 4 times longer than in the previous experiments. \dPVS{The results are indicated by the solid squares in Figure 9(a).} Interestingly, $\alpha_\mathrm{eff}$ reduces by approximately a factor of two as compared to the results obtained for shorter propagation lengths.  It is unlikely that the reduced absorption results from differences in propagation direction: in fact, previous studies have reported similar damping coefficients for LBAW propagation along the $\langle100\rangle$ and $\langle111\rangle$ directions [see, e.g., the results from Ref.~\cite{Cottam_JPCSSP7_2447_74} displayed in Fig.~\ref{Figure9}(b)]. Furthermore, the reduced absorption with longer propagation lengths also enables us to discard phonon diffraction as a loss mechanism. We rather assign the lower $\alpha_\mathrm{eff}$ to the reduced role of losses at interfaces and surfaces in the longer samples. 
The role of surface defects has been highlighted in recent studies of the absorption losses of acoustic resonators at low temperatures~\cite{Manenti_PRB93_41411_16,Hamoumi_PRL120_223601_18}. As an important consequence, the present results show that the acoustic absorption at low temperatures is limited by extrinsic (rather than by intrinsic) mechanisms, which can be minimized by improving the quality of the surfaces and interfaces.
 
\vspace*{-0.1cm}

\section{\label{Conclusions}Conclusions}

We have demonstrated the feasibility of piezoelectric generation of LBAWs in GaAs crystals with frequencies up to 20~GHz. The LBAWs were excited using BAWRs based on a thin ZnO layer produced by rf-sputtering as a piezoelectric medium. The BAWRs were fabricated using a simple process that combines the evaporation and sputtering through a shadow mask to define the lateral dimensions of the bottom contact and ZnO film, respectively, with the photolithographic fabrication of the top contact, which defines the active area of the structures. We showed that the structural properties of the ZnO layers, as well as the electro-acoustic conversion efficiency of the BAWRs, depend sensitively on the sputtering conditions of the ZnO films. Finally, we demonstrated that by reducing the thickness of the ZnO film one can increase the resonance frequency of the structures up to 19.6~GHz. 

The electro-acoustic properties of the BAWRs were probed by rf reflection measurements in the frequency and time domains. BAWRs deposited directly on GaAs substrates exhibit low ($30-40$~\%) electro-acoustic power conversion efficiency, which is mainly limited by the large impedance mismatch between the electric and acoustic loads. Significantly higher conversion efficiencies were achieved on structures fabricated on an acoustic DBR. By recording the acoustic echoes created by reflections at the sample boundaries we have developed a procedure to (i) measure the LBAW propagation losses with very high accuracy and (ii) determine the absolute electro-acoustic conversion efficiency into a LBAW.
The SHF BAWRs have enabled us to measure the acoustic absorption coefficient $\alpha_\mathrm{eff}$ with high accuracy in the previously unexplored frequency range from 1 to 20 GHz. The acoustic absorption coefficient measured at room temperature agrees very well with the predicted dependence expected from the  interaction with thermal phonons. Since the inverse of the LBAW lifetime increases superlinearly with frequency, the quality factor of acoustic resonators on GaAs reduces with frequency. As the temperature decreases, the absorption saturates at a frequency-dependent value, which appears to be determined by scattering at surfaces and interfaces. Due to the weak frequency dependence of the absorption at low temperature, the quality-factors of resonators should increase with frequency, thus becoming an  important consideration for the development of SHF resonators. Furthermore, the dependence of the acoustic absorption at low temperatures with the propagation length shows that the losses are not of  intrinsic character and can, thus, be minimized by proper design and control of the device interfaces. The present studies thus pave the way for the realization of high-quality nano-mechanical resonators operating in the SHF range as well as for use of these vibrations for the modulation and control of semiconductor nanostructures.

\paragraph{Acknowledgements}

We would like to thank B. Perrin, B. R\"{o}ben (PDI), M. Msall, and \aPVS{A. Hern\'{a}ndez-M\'{i}nguez (PDI)} for discussions and for a critical review of the manuscript. We also acknowledge the technical support from R. Baumann, S. Rauwerdink, and A. Tahraoui, which was essential for establishing the sample fabrication process. We acknowledge financial support from the German DFG (grant 359162958) and FAPESP (Brazil, process 2017/24311-6).




\bibliographystyle{apsrev4-2}
\bibliography{word}




\end{document}